\documentclass[structabstract]{aa}
\newcommand{\HII}{H\,{\scriptsize II}}

\newcommand{\CII}{[C\,{\scriptsize II}]}

\newcommand{\OI}{[O\,{\scriptsize I}]}

\usepackage{graphicx}
\usepackage{txfonts}
\usepackage{longtable}
\usepackage{natbib}
\bibpunct{(}{)}{;}{a}{}{,}

\begin{document}

\title{The ionizing source of the bipolar HII region S106: a close massive binary
\thanks{Based on observations obtained at the Observatorio del Teide, Canary Islands, Spain; and at the Centro Astron\'omico Hispano-Alem\'an, Calar Alto, Spain}
}
\author{F. Comer\'on\inst{1}
\and N. Schneider\inst{2,3}
\and A.A. Djupvik\inst{4}
\and C. Schnugg\inst{5}
}
 \institute{
  European Southern Observatory, Alonso de C\'ordova 3107, Vitacura, Santiago, Chile\\
  \email{fcomeron@eso.org}
  \and
  I. Physik Institut, University of Cologne, D-50937 Cologne, Germany
  \and
  LAB, CNRS, Universit\'e Bordeaux, F-33615 Pessac, France
  \and
  Nordic Optical Telescope, Apdo 474, E-38700 Santa Cruz de La Palma, Spain
  \and
  Visiting Scientist, European Southern Observatory, Alonso de C\'ordova 3107, Vitacura, Santiago, Chile\\
  }
%
%
\date{Received; accepted}
\abstract
{S106 is one of the best known bipolar HII regions, thoroughly studied and modelled at infrared, submillimeter and millimeter wavelengths, and it is one of the nearest examples of the late stages of massive star formation in which the newly formed star that ionizes it is still surrounded by vast amounts of gas and dust. However, little is known about its heavily obscured central source, S106IR.}
{The possible binarity of the central source is investigated, which is considered to be likely given the high binarity fraction among massive stars.}
{We have carried out visible and near-infrared photometric monitoring looking for short-term variability, with special interest in that related to the presence of a close binary companion to S106IR that may produce periodic eclipses or tidal distortion of the shape of the members of the system.}
{A periodic variability of S106IR in the $J$ band is found with a period of 5.0 days and an amplitude of $\simeq 0.1$~mag. The light curve displays a slow rise from minimum to maximum followed by a steep decrease, and can be well reproduced by a close binary system composed of two stars with different luminosity orbiting each other in an elliptical orbit of moderate eccentricity. S106IR also shows hints of short-term variability possibly related to accretion. We also report variability of four other stars previously classified as members of the S106 cluster, all of which are strong X-ray emitters.}
{The newly discovered close binarity of S106IR adds a new element to the modeling of the nebula and to the understanding of the dynamics of the gas around the ionizing source, which suggests that the components of the binary are accreting via a circumbinary disk. Binarity also helps to explain the apparent mismatch between the spectral type of the ionizing source inferred from the nebular spectrum and its high brightness at near-infrared wavelengths.}



\maketitle

\section{Introduction} \label{intro}

S106 is one of the most thoroughly studied \HII\ regions in our galactic vicinity (see \citet{Hodapp08} for a review). Images of S106 in the visible and infrared \citep[e.g.][]{Oasa06} show two distinct lobes where ionized gas flows in a cavity carved in the surrounding molecular cloud, as well as a cluster of mostly low-mass stars first reported by \citet{Hodapp91}. Mid-to far-infrared imaging \citep{Motte07,Simon12,Adams15} show that the dark lane separating the two lobes of S106 is a very high column density filamentary feature associated to the molecular cloud. High angular resolution observations at infrared and radio wavelengths also have shown a small, nearly edge-on disk \citep{Hoare94,Gibb07}. Molecular line maps \citep[][and Schneider et al. 2017, submitted.]{Schneider02} show a complex velocity structure in the proximity of S106IR, including high-density clumps and small-scale flows of gas.

Despite the recent improvements in modeling, little is still known about {\sl the exciting source} S106IR. It is bright at infrared wavelengths ($J=10.36$, $H=7.73$, $K_S=5.87$ in 2MASS), but heavily obscured in the visible, where it was first detected by \citet{Pipher76}. It was then further studied by \citet{Eiroa79} and \citet{Gehrz82}. Its characteristics have been indirectly inferred from its infrared luminosity, the excitation of the \HII\ region, and the properties of the photodissociated gas \citep{Hodapp08,Vandenancker00,Schneider03,Simon12,Stock15}. All observations were consistent with a very young massive stellar object with estimated spectral types between O7 and B1.

The distance to the nebula is $1.3 \pm 0.1$~kpc \citep{Xu13}, determined from maser parallax measurements. The adoption of that distance yields a luminosity somewhat exceeding that expected for a zero-age main sequence star in the estimated range of spectral types, perhaps hinting at the existence of a companion contributing to the observed flux. In addition, recent high angular THz spectroscopy of the \OI\ 63 $\mu$m and \CII\ 158 $\mu$m lines with SOFIA\footnote{Stratospheric Observatory for Far-Infrared Astronomy} (Schneider et al. 2017, submitted) show a very complex emission distribution that can not be explained by a simple star-disc system. Besides these indirect indications, S106IR may be expected to be a massive binary based on purely statistical grounds, as most massive stars are born in multiple systems \cite[e.g.][]{Chini12,Aldoretta15}. However, no direct evidence on the multiplicity of S106IR has been reported thus far, and the indirect hints provide essentially no information on the characteristics of possible companions.

In this paper we report the results of two photometric monitoring campaigns in the near-infrared and the visible that reveal a definite periodic variability signal demonstrating {\sl the binary nature of S106IR}, and we derive some constraints of the characteristics of the companion and the orbital parameters of the system. We also find evidence of short-term, low-amplitude variability that we tentatively associate to accretion activity. Furthermore, we report on high angular resolution imaging of S106IR that excludes the presence of massive wide companions. Finally, we present light curves of four members of the S106 cluster, previously recognized thanks to their X-ray emission, that display clear variability.

\section{Observations and data reduction\label{observations}}

\subsection{Near-infrared monitoring\label{cain}}

Near-infrared monitoring of the S106 field was carried out on 14 consecutive nights between 23 July and 6 August
2017 using the CAIN infrared camera at the 1.5m Carlos S\'anchez telescope of the Observatory of the Teide (Canary Islands, Spain) through the wide-field optics yielding a field of view of $4'7 \times 4'7$ on the $256 \times 256$ pixel NICMOS3 detector. The observations were taken through the $J$ filter using integration times of $5$~s to keep detector counts within the linear regime. Each individual observation consisted of a raster of 9 individual pointings on a square grid of $40'' \times 40''$. A sky frame was obtained for each observation by median filtering the 9 exposures without correcting for the telescope offsets, and it was subtracted from each of the exposures, which were then divided by a dome flat field calibration frame. The sky-subtracted, flat-fielded frames were then shifted to correct for the telescope offsets and stacked together. Three observations were obtained before moving to other targets unrelated to this program, and the sequence was then carried out again leaving typically an interval of 30 minutes with respect to the previous one. The scheduling of the observations in late July/early August, when S106 culminates near local midnight, allowed us to observe from twilight to twilight thus obtaining a total of nearly 200 sequences spread over the 14 nights.

Aperture photometry was carried out and calibrated using as a flux reference the 2MASS magnitudes of unsaturated
and isolated stars in the same field. One magnitude measurement of each detected source for each sequence was obtained by median averaging the magnitudes measured from each of the three observations composing it. Light curves were thus obtained for all unsaturated stars in the field, and the absence of systematic effects that might result in spurious variability signals was thus verified.

\subsection{Optical monitoring\label{camelot}}

Observations of S106 were obtained on 5 consecutive nights between 15 and 20 September 2016 using the CAMELOT camera at the 0.80~m IAC80 telescope, also at the Observatory of El Teide. A filter reproducing the Gunn $z'$ passband was used given the very high extinction of S106IR at visible wavelengths. The camera has a detector providing a $10'4 \times 10'4$ field of view with a pixel scale of $0''305$ that adequately oversamples the point-spread function in the seeing conditions under which our observations were obtained. We took series of 3 exposures of 60~s each, then moving to other targets unrelated to this program before returning to the field of S106 and repeating the sequence. The longer interval between sequences, and the scheduling of our observations in September when S106 appears at an elevation below $30^\circ$ in the last two hours before dawn reduced the span of our observations to approximately 6.5 hours per night, thus allowing us to obtain 210 individual observations over the 5 nights.

We chose the PSF-fitting method to obtain the photometry in the optical observations, given the oversampling of the stellar images and the need to minimize contamination by nearby bright nebulosity under variable seeing conditions, which could be much better carried out in this way than by means of aperture photometry. Flux calibration was done by using as a reference the SDSS $z'$ magnitudes of unsaturated and isolated stars in the field.

\subsection{Lucky imaging\label{astralux}}

High angular resolution images of S106IR and its immediate surroundings were obtained on the night of 27 August 2015 using AstraLux \citep{Hormuth08}, the lucky imaging camera at the 2.2~m telescope of the Calar Alto Observatory (Spain), using a Gunn $z'$ filter. In combination with the 2.2~m telescope AstraLux yields a pixel scale of $0''047$ on the $512 \times 512$ detector. Given the faint $z'$ magnitude of S106IR we used the maximum exposure time, 200~ms, that still allowed us to 'freeze' atmospheric turbulence. To optimize the possibilities of sampling the best turbulence conditions occurring on that night we obtained 14 sequences of observations, each consisting of 10,000 individual images, so as to be able to select the best ones at the processing stage.

The combination of pixel scale and exposure time results in very low illumination levels per pixel for a source as faint ($z' = 15$) as S106IR, which prevented the use of the instrument pipeline that automatically selects the best images of the target in each frame and determines an accurate centroid that is used to shift and add them. Instead, we first binned the frames in order to obtain lower resolution images where S106IR appeared with enough signal to allow a meaningful determination of the centroid, and applied the shifts determined in this way to the unbinned images before stacking them. The binned images were also used to obtain an estimate of the full-width at half-maximum of the image of S106IR in order to select those obtained under the best turbulence conditions. We experimented with various binning factors in order to optimize the trade-off between high resolution and sufficient signal for accurate centroiding, finding that $2 \times 2$ binning offered the best compromise. We also experimented with changing the percentage of selected images in each sequence, and decided in this way to retain, shift and stack the 1\% best images of each. Not surprisingly, the best results were obtained with the sequence obtained at the lowest zenith distance.

\section{Results\label{results}}

\subsection{The light curve of S106IR\label{lightcurve}}

\begin{figure}[ht]
\begin{center}
\hspace{-0.5cm}
\includegraphics [width=8.5cm, angle={0}]{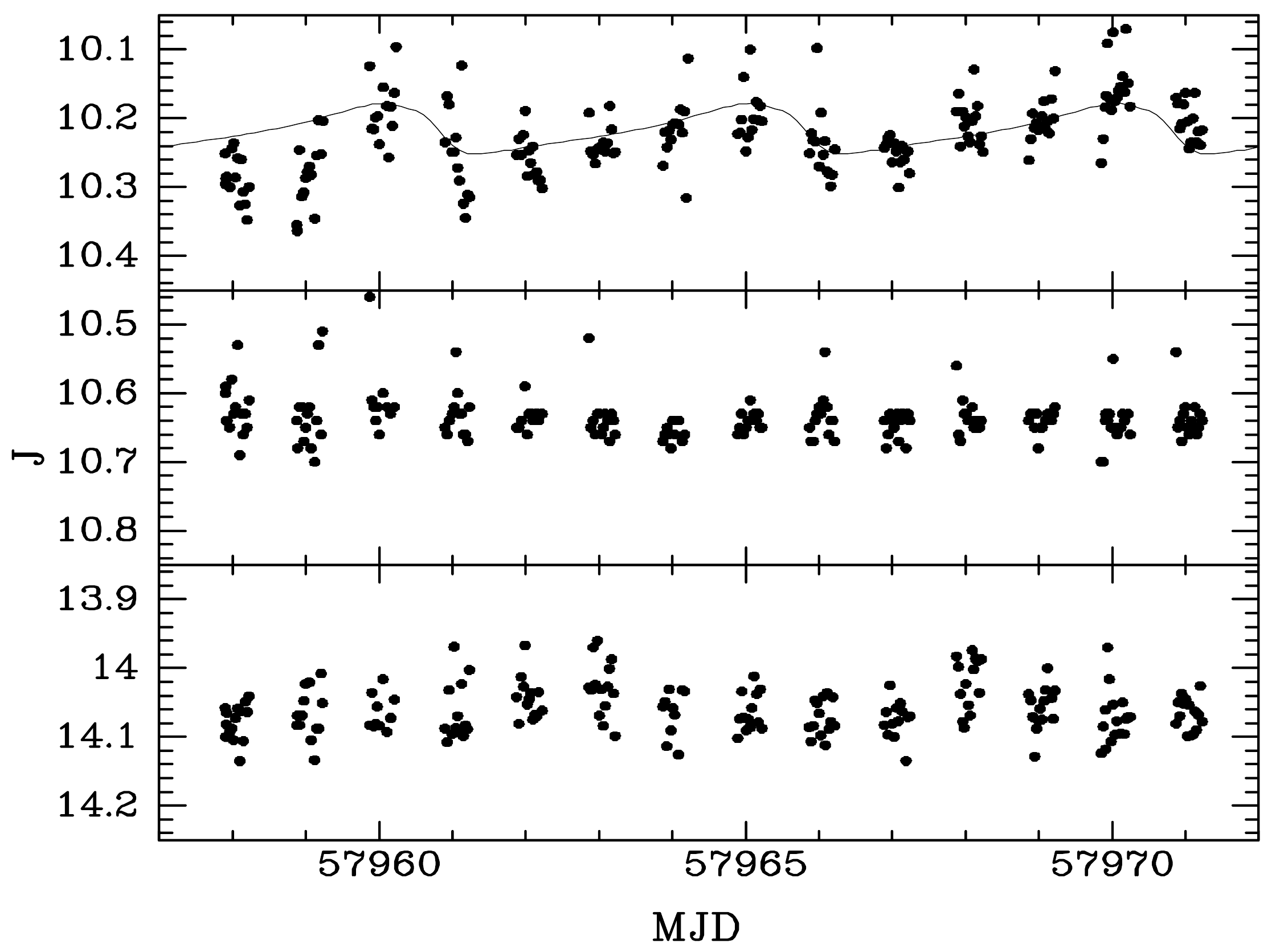}
\caption []{$J$-band light curve of S106IR. The solid line is the fit to the $J$ light curve obtained with the Wilson-Devinney program as described in Section~\ref{system}. The measurements for two comparison stars in the field are shown for reference. The middle panel shows the light curve of 2MASS J20272671+3721149, a star of magnitude similar to S106IR. The smaller scatter s due to its position projected on an area devoid of nebulosity. The lower panel shows the light curve of 2MASS J20273393+3722349, a fainter star whose measurements are also unaffected by nebulosity.}
\label{lightcurveJcomp}
\end{center}
\end{figure}

  Figure~\ref{lightcurveJcomp} shows the $J$- and $K_S$-band light curve of S106IR, with an obvious oscillation of the
former of approximately 0.08~mag amplitude and a period of approximately 5 days. The variability curve is also clearly asymmetric, with a steady rise to the maximum followed by a rapid fall to a minimum over a time span of one day or less. In fact, Figure~\ref{lightcurveJcomp} suggests that we may have sampled most of the drop from peak to minimum brightness on our fourth night of observations.

  Several types of pulsating variables are known among the otherwise generally stable B stars \citep{Pamyatnykh99},
but none occupies the period-amplitude region where S106IR appears. $\beta$ Cep variables \citep{Stankov05} are early-type B stars with typical periods below one day and amplitudes of few hundredths of magnitude. Slowly-pulsating B stars \citep{Waelkens91} have in general later spectral types and their main periods do not exceed $\sim 2$~days, being generally shorter. The class most resembling the variability characteristics of S106IR is that of $\alpha$~Cyg variables, but those are evolved massive stars on their way away from the main sequence \citep{Saio13} and thus very unlikely to be related to an extremely young object like S106IR. Finally, the lack of a convective envelope among B stars rules out magnetically driven starspots as the cause for variability. We are thus confident that intrinsic causes can be excluded as being responsible for the variability that we observe in S106IR, leaving binarity as the most likely candidate. This confidence is reinforced by the existence of a realistic set of system parameters that provide a good fit to the observations, as discussed in the next Section. We note that the $J$ light curve in Figure~\ref{lightcurveJcomp} shows hints of a long-term trend superimposed on
the periodic variation, consisting of a steady increase in brightness at a rate of $\sim 0.03$~mag per period. The cause of this slow increase is unclear, and we speculate that it might be due to varying foreground extinction toward the binary caused by transiting structures in the obscuring dust of the bar separating the two lobes of the nebula.

\subsection{Parameters of the S106IR binary system\label{system}}

  The period of variability of 5.0 days that we derive is well into the range of close, non-contact binaries where
most massive binary systems reside. The small distances between components imply strong tidal effects that can severely distort the shape of the components of the system, and cause their mutual illumination, making their flux depend on the aspect angle under which we observe them and thus causing periodic variability directly linked to the orbital motion. The asymmetric shape of the light curve is moreover an indication that the orbit is non-circular, and the lack of obvious eclipses indicates that the line-of-sight forms a significant angle with respect to the orbital plane.

  We have carried out an extensive series of tests with the most recent version of the widely used Wilson-Devinney
program \citep{Wilson71,Wilson79} to better constrain the parameters of the S106IR system. We use as an imposed constraint the existence of at least one O9V star with a temperature $T_{\rm eff} = 32,500$~K \citep{Pecaut13} to be consistent with the derived ionization properties of the \HII\ region. We have obtained no acceptable fit to the light curve with similar-mass components regardless of the eccentricity, inclination and argument of periastron of the orbit. This is due to the fact that the complementary aspect angles of the two components along the orbit combine to compensate their respective flux variations, except at very specific positions of the orbit around the periastron, producing an essentially flat light curve over most of the period.

\begin{figure}[ht]
\begin{center}
\hspace{-0.5cm}
\includegraphics [width=8.5cm, angle={0}]{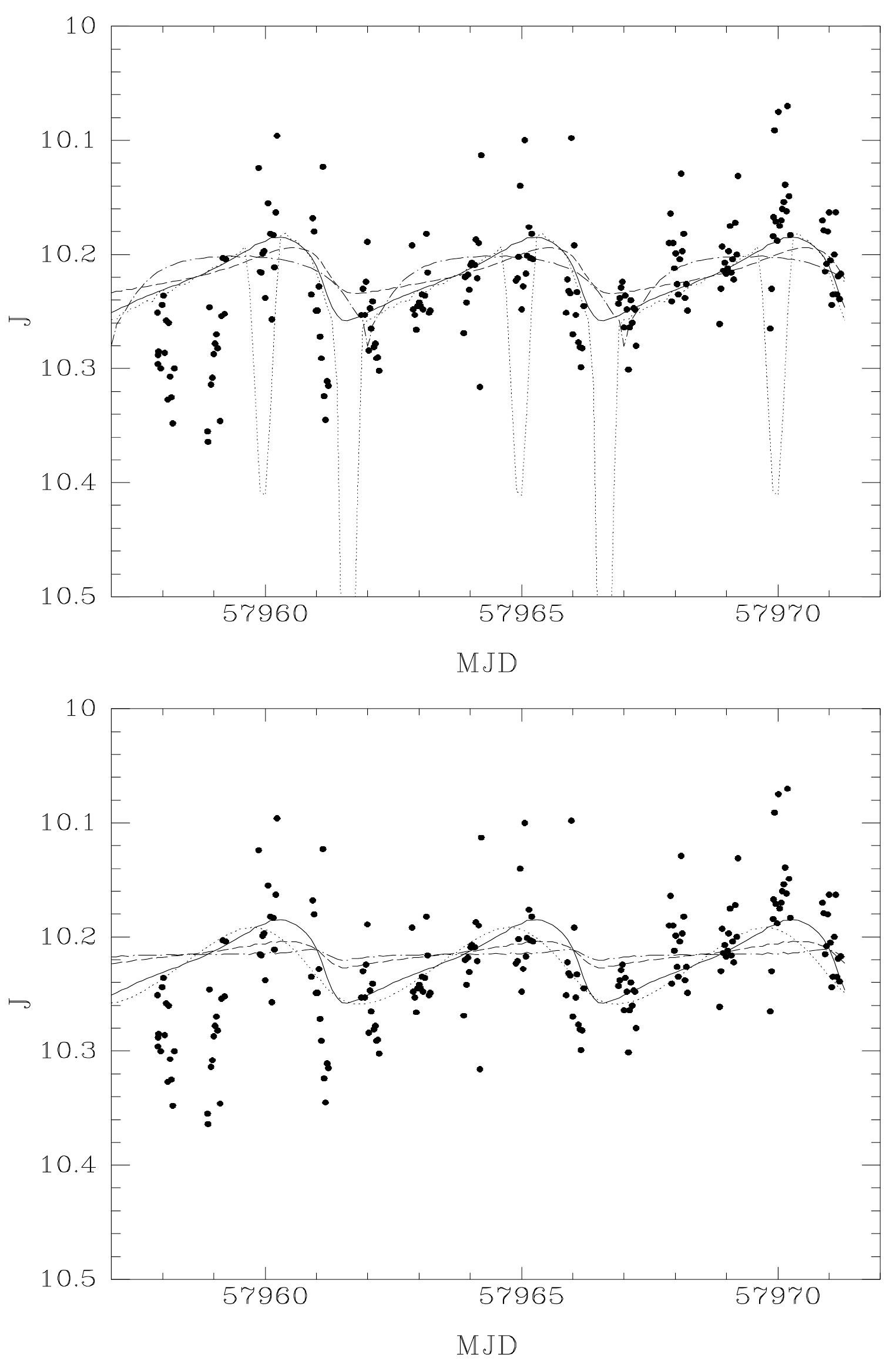}
\caption []{Different Wilson-Devinney fits to the $J$-band light curve of S106IR. {\it Top panel:} Solid line: best fit model with the parameters given in Table~\ref{params}. Dashed line: same as the best-fit model but with an inclination $i=30^\circ$, closer to a pole-on view of the orbit. Dotted line: same as the best-fit model but with an inclination $i=90^\circ$, corresponding to an edge-on view of the orbit in which the two components periodically eclipse each other.  Dot-dashed line: same as the best-fit model with with an argument of the periastron of $90^\circ$. {\it Bottom panel:} Solid line: best-fit model. Dashed line: same as the best-fit model but with a secondary of $M_{\rm bol} -1$. Dotted line: best-fit model with eccentricity $e = 0.1$. Dot-dashed: same as best-fit model but with a secondary with $T_{\rm eff} = 20000$~K.}
\label{lightcurvecomp}

\end{center}
\end{figure}

  The shape of the light curve is nevertheless well reproduced with two unequal stars, so that the distortion-induced
variability of the primary is not offset by that of the secondary. The steady rise in magnitude that we observe can be well reproduced by an elliptical orbit in which the periastron takes place around mid way between the inferior conjunction of the secondary and the quadrature. Other possible locations of the periastron reproduce the amplitude of variability, but not the shape of the rise to maximum. To estimate the properties of the components we have assumed that the O9V primary, being substantially more massive than the secondary, has the same absolute bolometric magnitude $M_{\rm bol}$ as a single main-sequence star of the same spectral type, determining in this way the value of the potential $\Omega_1$ on the surface of the primary. We have then tried to reproduce the light curve for several choices of the secondary ranging from B1V to A0V spectral types, taking its $T_{\rm eff}$ and mass from \citet{Pecaut13} and varying the value of the potential on the surface $\Omega_2$, thus varying its radius and the amplitude of the light curve. We obtain good fits for types B5 and later, always by choosing $\Omega_2$ values that yield an absolute magnitude $M_{\rm bol}$ substantially brighter than the main sequence value for the same spectral type, thus indicating that the tidal effect of the primary causes the star to have a large radius than in the main sequence. The difference between the best-fit $M_{\rm bol}$ and the main-sequence value for the same spectral type passes through a distinct minimum at spectral type B8 ($T_{\rm eff} = 12,500$~K), which we adopt as the preferred type. Even so, the difference in $M_{\rm bol}$ reaches $-1.46$~mag with respect to the main sequence value, implying a radius a factor of 2 larger than that of a single main sequence star of the same type. The contribution of the companion to the combined $J-$band flux of the system is approximately 20\%.

  Assuming main sequence masses of $19.6$~M$_\odot$ for the O9 primary and $3.4$~M$_\odot$ for the B8 secondary (mass
ratio $q = 0.17$), the 5.0~day period implies a semimajor axis of the orbit of $0.17$~AU, or $35$~R$_\odot$. The eccentricity is rather loosely constrained, as a relatively wide range of values between $e = 0.1$ and $e = 0.3$ produce similarly good fits to the observed light curve. Likewise, it is not possible to give a precise value of the inclination of the orbit from the fit: $i < 60^\circ$ (where $i = 90^\circ$ would be edge-on) is well established thanks to the absence of eclipses, but the effect of decreasing inclination on the decrease of the amplitude of the light curve is slow. However, given the orientation of the general symmetry axis of S106 near the plane of the sky it seems safe to assume that the inclination of the orbit must be close to that upper limit. Table~\ref{params} summarizes the adopted and derived parameters of the S106IR binary system, and Figure~\ref{lightcurveJcomp} shows the Wilson-Devinney fit to the light curve.  To illustrate the influence on the light curve of the various parameters characterizing the system, Figure~\ref{lightcurvecomp} gives sample fits obtained by changing some of them with respect to the best-fit values listed in Table~\ref{params}.

\begin{table}[t]
\caption{S106IR binary system parameters}
\begin{tabular}{lc}
\hline\\
Period:         & $5.0 (\pm 0.1)$ days \\
Semimajor axis: & $0.17$ AU \\
Eccentricity:   & $0.2 (\pm 0.1)$ \\
Inclination:    & $\sim 60^\circ$ \\
Argument of periastron: & $30^\circ (\pm 10^\circ)$ \\
Mass ratio:     & $\sim 0.17$ \\
Primary star (adopted): & \\
\quad Spectral type: & O9 \\
\quad Temperature: & $32,500$ K \\
\quad Bolometric magnitude: & -7.2 \\
Secondary star (estimated): & \\
\quad Spectral type: & B8 \\
\quad Temperature: & $12,500$ K \\
\quad Bolometric magnitude: & -2.4 \\
\hline\\
\end{tabular}
\label{params}
\end{table}

\subsection{Short-term variability of S106IR\label{accretion}}

\begin{figure}[ht]
\begin{center}
\hspace{-0.5cm}
\includegraphics [width=8.5cm, angle={0}]{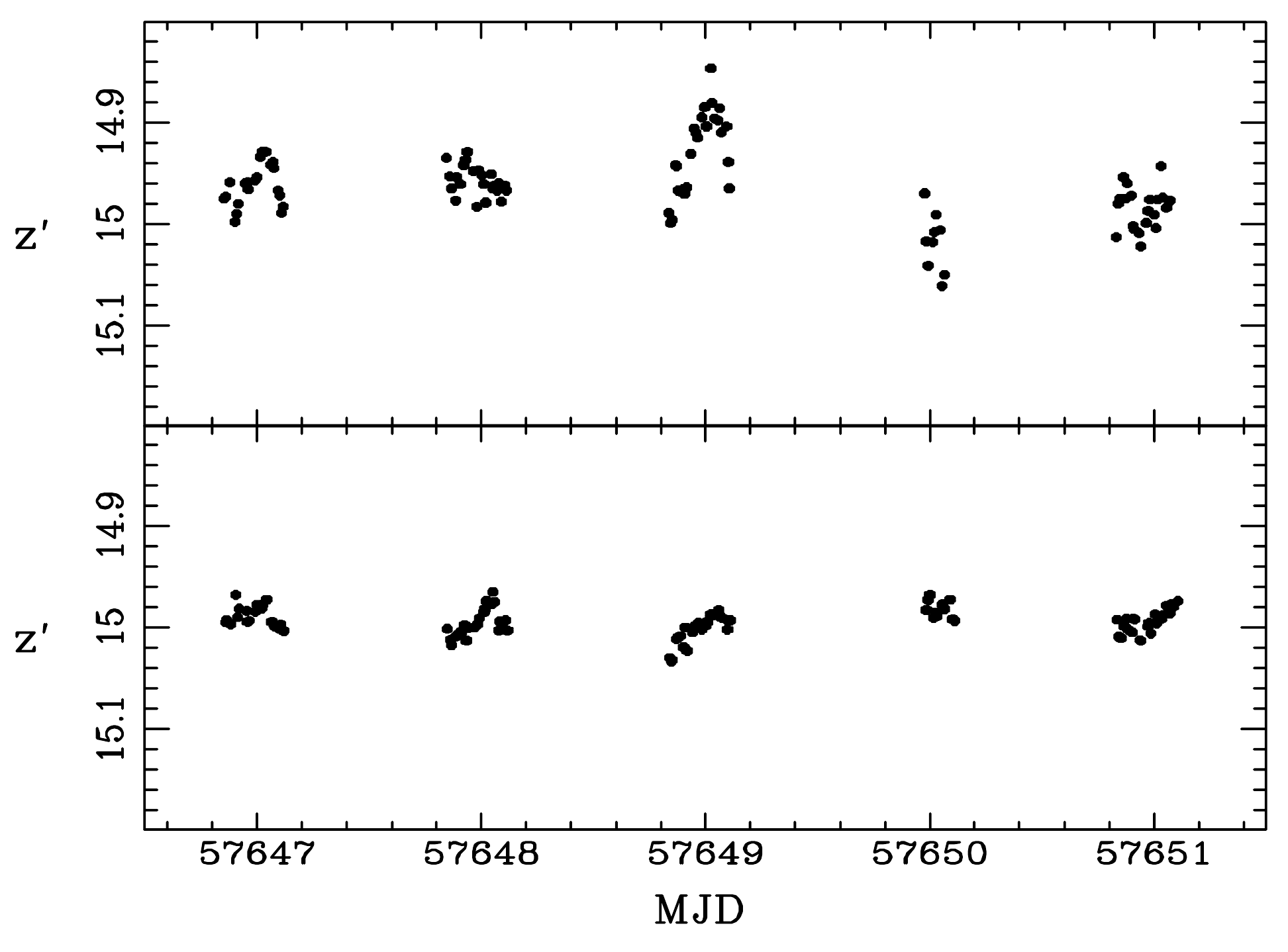}
\caption []{Gunn $z'$-band light curve of S106IR over a time span of five days (top panel). Note the rather large short-term variations on some of the nights, with a period of hours that is much shorter than the main period of variability of 5~days.
The measurements of a comparison star in the field of very similar brightness in $z'$, 2MASS J20273205+3723571, is shown for reference.}
\label{lightcurvez}
\end{center}
\end{figure}

Our light curve of S106IR in the visible (Figure~\ref{lightcurvez}) covers a time span of the order of the variability period in the infrared. The signature of the 5-days periodicity is far less obvious, but it can still be seen that the nightly-averaged flux peaks on the third day of our optical observations and clearly drops between the third and the fourth nights by $0.05 - 0.10$ magnitudes, thus reproducing the periodic sharp drop in flux that we observed three times during the two weeks spanned by our $J$-band observations. The average $z'$-band flux in the last two days is below that measured in the first three, thus suggesting that the $z'-$ band light curve closely tracks the variations seen in the $J-$band.

The superior quality of the photometry in $z'$ also reveals what appear to be short-term variations with a characteristic timescale of hours. This is probably seen in the photometric evolution on the first night, where the rise and decrease in magnitude has an amplitude of $0.08$~mag, above the $\pm 0.03$~mag uncertainty of the individual measurements, but especially on the third night, in which the measurements trace a rise and decrease of a full $0.12$~mag over the six-hours spanned by the observations on that night. In the absence of simultaneous confirming  spectroscopic observations we tentatively associate such short-term variations to accretion bursts on the surface of one of the two members of the binary, and we note that the episode with the largest amplitude is detected around the peak of the light curve concurrently with the periastron passage. This may not be coincidental and may be a phenomenon similar to that of pulsed accretion observed in other young stellar objects that are part of close binary systems \citep{Tofflemire17}, even with more extreme bursts \citep{Muzerolle13}. The conclusion is nevertheless speculative as it is based on the photometric behavior on a single night. Future observations, especially long-term photometric monitoring, may confirm if the short-term bursts systematically happen near periastron passages, and if they are indeed associated with accretion.

\subsection{No distant massive companions to S106IR\label{distant}}

\begin{figure*}[ht]
\begin{center}
\hspace{-0.5cm}
\includegraphics [width=14cm, angle={0}]{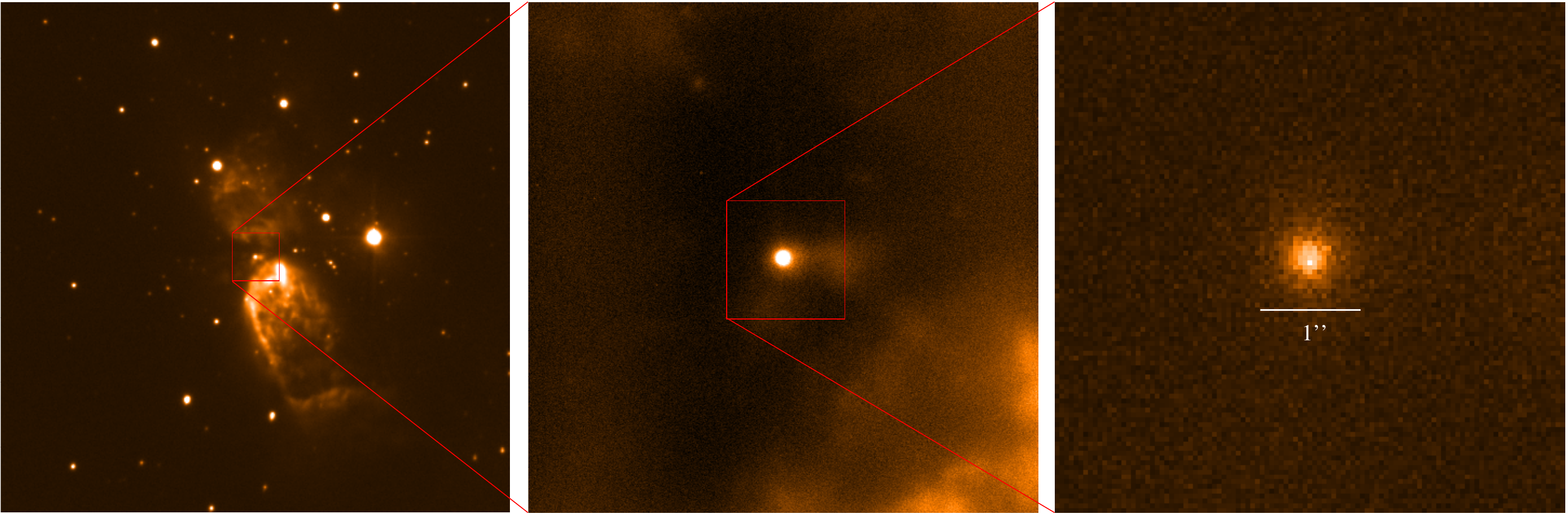}
\caption []{A zoom onto S106IR in the $z'$ band, showing a general view of the nebula obtained with the IAC80 telescope (Section~\ref{camelot}, left panel); the field of view covered by the AstraLux observations (middle panel); and a detailed view of the best AstraLux image of S106IR (right panel).}
\label{lucky}
\end{center}
\end{figure*}

Given the faintness of S106IR at visible/red wavelengths and the absence of any sufficiently bright stars at a close enough angular distance, our lucky imaging observations cannot exploit the full capabilities of the technique to provide diffraction-limited images. However, the results obtained when processing the data in the way described in Section~\ref{astralux} deliver a quality vastly exceeding that of seeing-limited observations. The point-spread function of the best image of S106IR, presented in the right panel of Figure~\ref{lucky}, displays the typical shape of images obtained with this technique, with a sharp core having a full-width at half-maximum of $0''24$ and extended wings. We note that deep near-infrared observations of the region were obtained by \citet{Oasa06} with excellent image quality (full-width at half-maximum $0''35$) but S106IR is saturated in them, thus limiting the capacity to detect objects at subarcsecond distances. Likewise, the existing infrared images obtained with the Wide-Field Camera 3 onboard the Hubble Space Telescope also show S106IR substantially saturated. Our lucky imaging observations are thus the best suited observations to date to search for possible massive companions around S106IR in the $\sim 0.2''-1''0$ range.

\begin{figure}[ht]
\begin{center}
\hspace{-0.5cm}
\includegraphics [width=8.5cm, angle={0}]{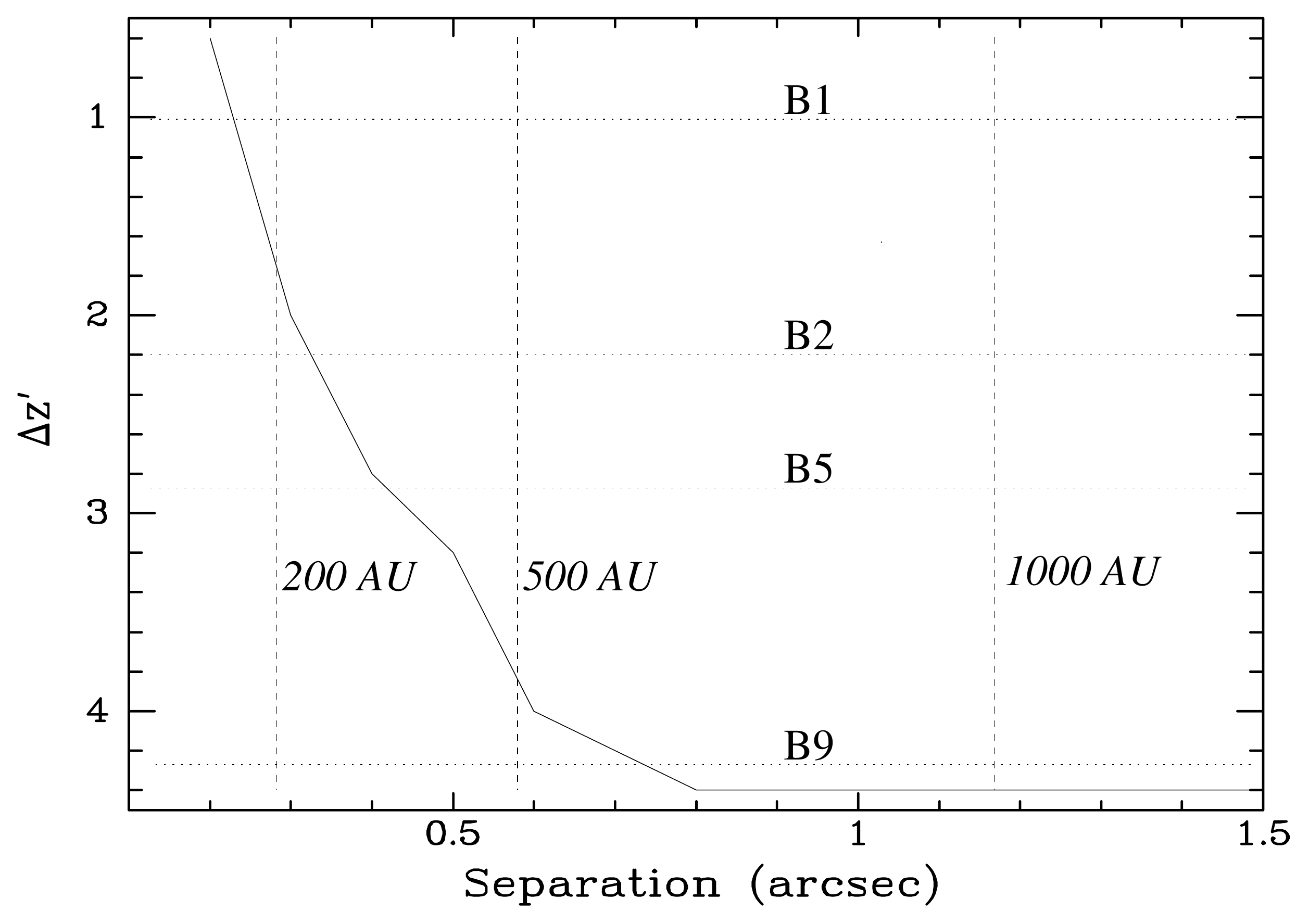}
\caption []{Detectability limits of possible companions to S106IR, obtained as described in Section~\ref{distant}. The solid line marks the closest separation at which a star with a given difference of $z'$ magnitude with respect to that of S106IR, $\Delta z'$, and obscured by the same amount of foreground extinction would still be distinctly detectable at the $3 \sigma$ level. Dotted lines show the spectral type corresponding to companions with different magnitude differences, and dashed lines indicated the projected distances corresponding to selected separations at
the distance of $1.3$~kpc.}
\label{detectability}
\end{center}
\end{figure}

Figure~\ref{lucky} shows at a glance that no other sources are found near S106IR, thus ruling out the possible
existence of distant companions detectable down to our sensitivity limits. To be more quantitative in ruling out companions we carried out simulations in which the image of S106IR was isolated, scaled to an arbitrary flux, and added at a given distance and position angle of the actual source to simulate a companion. The resulting image was analyzed to find the flux yielding a $3\sigma$ detection of the simulated companion at various distances of the central star. Results are displayed in Figure~\ref{detectability}, where the greatest $z'$ magnitude difference of the companion with respect to the central star still allowing its detection is given as a function of the angular distance between both. The results are also expressed in terms of projected distance to the primary and of the spectral type of the companion, assuming the primary to have a O9 spectral type and both the primary and the companion to be obscured by the same amount of foreground extinction. Our results thus safely rule out any B-type companions more distant than $\simeq 600$~AU of projected distance from the primary, reducing this limit to $\simeq 200$~AU for possible early B-type companions.

\section{Discussion\label{discussion}}

  The observations presented here provide new insights into the ionizing source of S106 and its circumstellar surroundings. The semimajor axis of the orbit of
the binary is smaller than the inner edge of the disk traced by CO bandhead emission in the $K$ band, $\simeq 0.3$~AU, thus suggesting that the molecular gas studied by \citet{Murakawa13} traces a circumbinary disk, and that accretion onto the components of the binary system takes place directly from it. The mass of the central system derived by \citet{Murakawa13} from the kinematics of CO, $22 \pm 5$~M$_\odot$, is too low for a binary system composed by two late O-type or very early B-type stars of similar mass, being instead consistent with our results that favor a system composed by two stars with a rather small mass ratio $q$ (see Sect.~\ref{system} and Table~\ref{params}). An intriguing discrepancy between our results and those of \citet{Murakawa13} is the inclination much closer to edge-on, $i = 83^\circ$, found by the latter, which would unavoidably result in deep eclipses if applied also to the orbital plane of a system with a period as short as the one that we measure. If the inclination derived by \citet{Murakawa13} were confirmed it might imply a misalignment between the disk plane and the orbital plane of the binary system. On the other hand, a lower (i.e. farther from edge-on) inclination of the disk would raise their dynamical mass estimate of the central system to $M = 29 \pm 6.5$~M$_\odot$, still consistent with our own estimate of $M \simeq 23$~M$_\odot$.

  Further away from S106IR, the complex environment of the central region of S106 has been the subject of detailed
study, both observationally and in modelling, for over two decades. Recent observations in the millimeter and infrared (Schneider et al. 2017, submitted) suggest that the dark lane separating the two lobes of the nebula in visible and infrared images is a filamentary structure essentially tracing a flow that has its end at S106IR. Further monitoring of the short-term variability may verify its proposed link with accretion onto the binary system and make possible estimates of the accretion rate to be compared with the parameters derived for the flow, thus yielding insights on the accretion efficiency. S106 thus offers a rare possibility of studying in detail the late stages of the formation of massive close binary systems and their circumstellar environments, both immediate and extended.

\begin{acknowledgements}

  We are very grateful to the staff of the Observatorio del Teide providing excellent support to the operation of the
IAC80 and Carlos S\'anchez telescopes, and in particular to \'Alex Oscoz for a generous allocation of time to this project that proved to be essential to its success. We are also equally thankful to the staff of Calar Alto observatory for their support during our AstraLux observing run. The constructive remarks by an anonymous referee helped to improve the presentation of our results. NS acknowledges support by the French ANR and the German DFG through the project "GENESIS" (ANR-16-CE92-0035-01/DFG1591/2-1). This publication makes use of data products from the Two Micron All Sky Survey, which is a joint project of the University of Massachusetts and the Infrared Processing and Analysis Center/California Institute of Technology, funded by the National Aeronautics and Space Administration and the National Science Foundation.

\end{acknowledgements}

\bibliographystyle{aa} 
\bibliography{s106ir_cit}

\begin{appendix}

\section{Other variable stars in the S106 cluster}

\begin{figure}[ht]
\begin{center}
\hspace{-0.5cm}
\includegraphics [width=8.5cm, angle={0}]{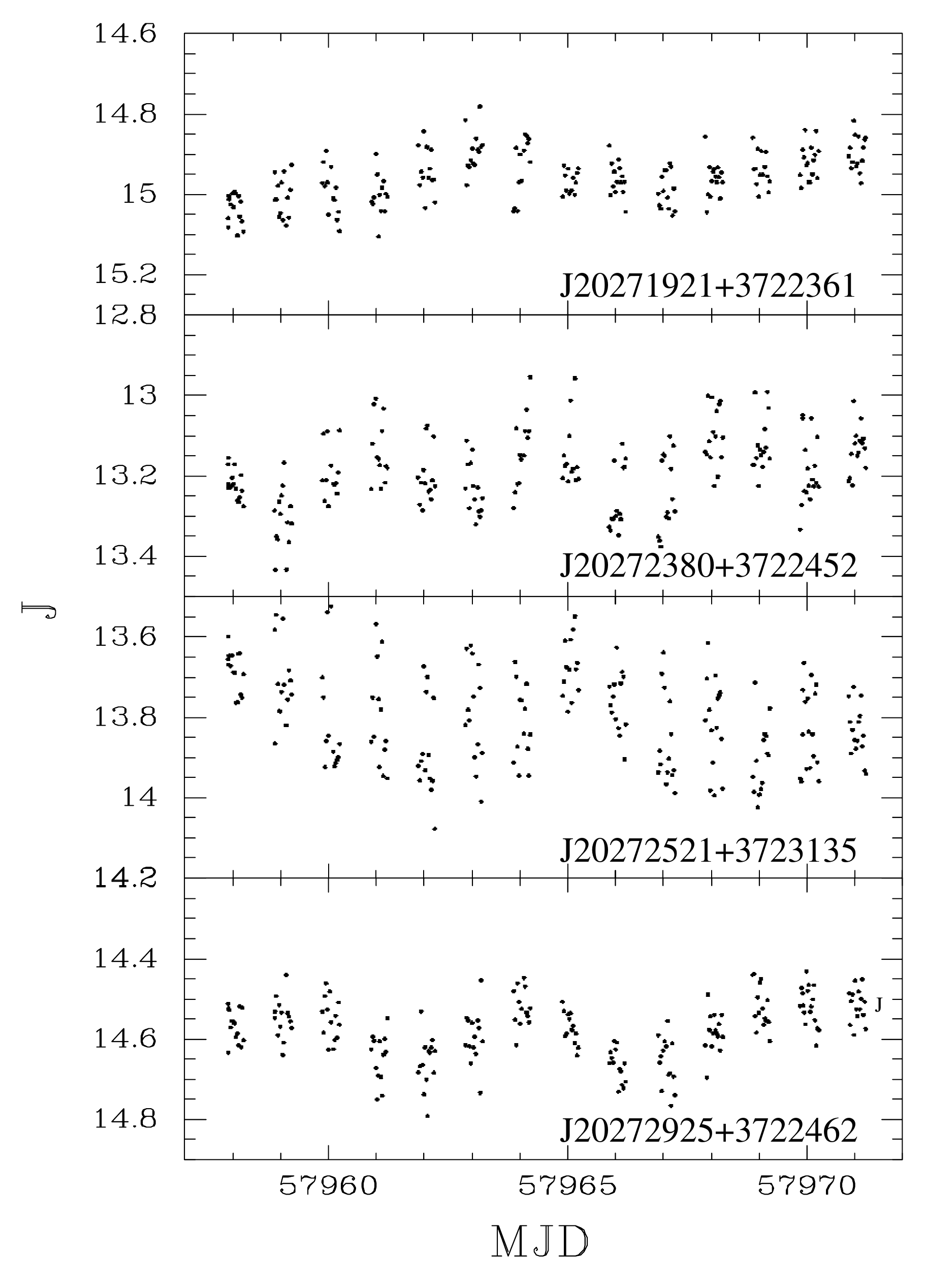}
\caption []{Light curves of the four additional variable stars identified in our observations.}
\label{othervars_J}
\end{center}
\end{figure}

\begin{figure}[ht]
\begin{center}
\hspace{-0.5cm}
\includegraphics [width=8.5cm, angle={0}]{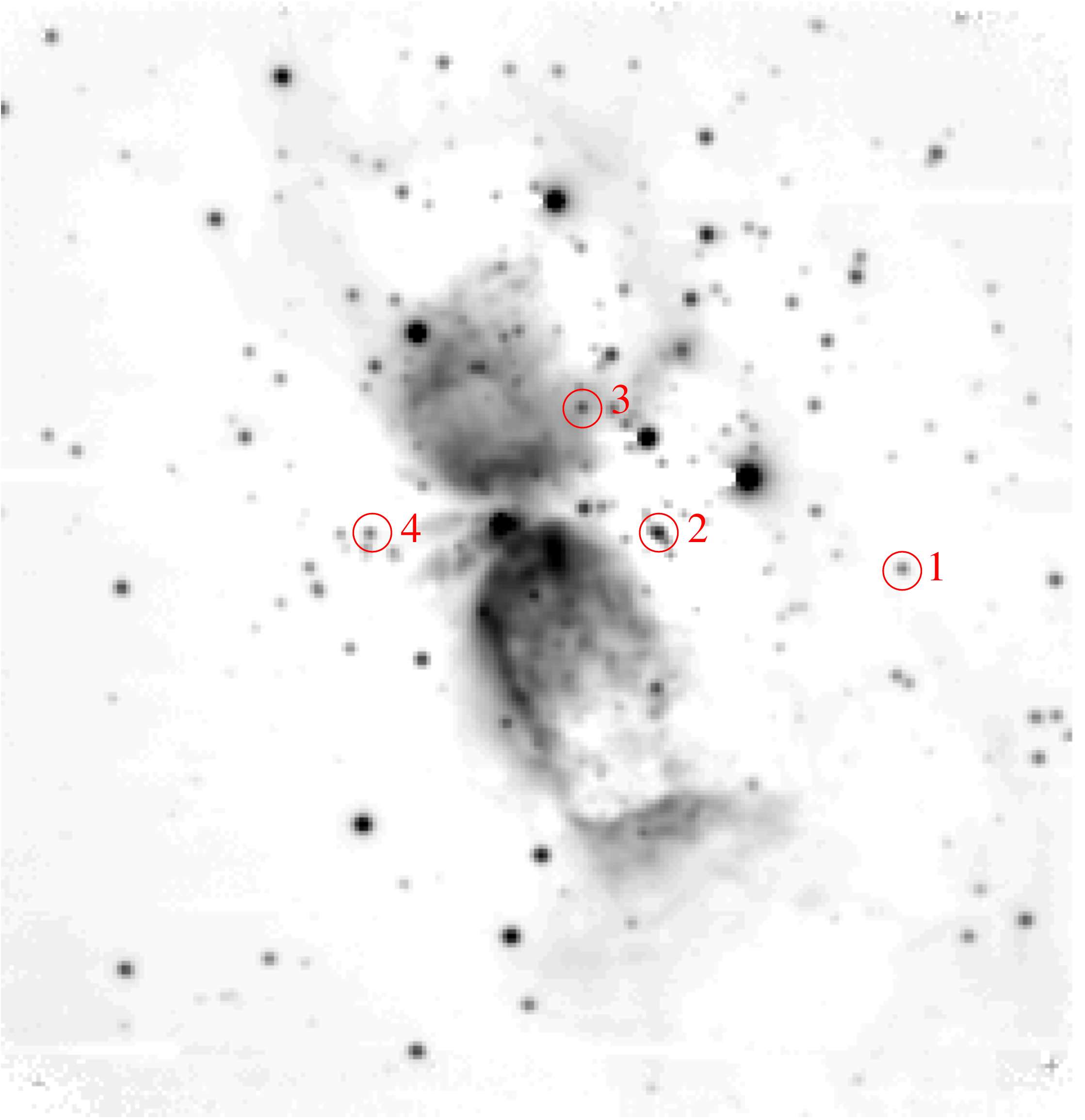}
\caption []{Location of the four additional variable stars identified in our observations on a $J-$band image of the field around S106 obtained by coadding all our observations. North is to the top and East to the left}
\label{othervars_map}
\end{center}
\end{figure}

\begin{table*}[t]
\caption{Other variable stars in the S106 field}
\begin{tabular}{llccccccc}
\hline
Nr. & 2MASS name          & $J^a$  & $H^a$  & $K_S^a$ & Mass$^b$    & $A_V$ & $\log L_X^c$ & $\log (L_X / L_{\rm bol}$)\\
    &                     &        &        &         & (M$_\odot$) & (mag) & (L$_\odot$)  & \\
\hline
1   & J20271921+3722361 & 15.006 & 13.534 & 12.938  & 0.3         & 3.6   & 30.8         & -2.6 \\
2   & J20272380+3722452 & 13.208 & 12.069 & 11.512  & 1.1         & 3.1   & 31.2         & -3.0 \\
3   & J20272521+3723135 & 14.302 & 12.615 & 11.715  & 1.0         & 6.0   & 31.6         & -2.5 \\
4   & J20272925+3722462 & 14.837 & 13.072 & 12.019  & 0.6         & 5.7   & 31.5         & -2.3 \\
\hline\\
\end{tabular}
\\Notes:\\
$^a$: Near-infrared magnitudes from \citet{Oasa06}\\
$^b$: Mass estimated from $JH$ photometry using evolutionary tracks from \citet{Baraffe15} for an age of 0.5~Myr\\
$^c$: X-ray luminosity from \citet{Giardino04} corrected for our adopted distance of 1.3~kpc\\
\label{othervars_pars}
\end{table*}

  The observations presented here yield as a byproduct the first monitoring study of variability of the S106 infrared cluster. We have obtained the light
curves of all the stars in the field detected in the $J-$band images. In addition to the reported results for S106IR, we find four other stars for which significant changes in the $\Delta J \simeq 0.1$~mag range take place. Their light curves are shown in Figure~\ref{othervars_J}, and their location in the field of S106 are shown in Figure~\ref{othervars_map}.

  The most regular behaviour is displayed by J20271921+3722361, with a period of $\simeq 8$~days and a close similarity between the almost two cycles covered
by the span of our observations. J20272380+3722452 seems to have more irregular variations in which a period of $\simeq 4$~days can be easily identified, and no clear periodicity is seen in the light curves of J20272521+3723135 (whose photometry is less accurate due to its location projected against bright nebulosity) and J20272925+3722462.

  To derive intrinsic properties of these stars we have taken their $JHK_S$ magnitudes from \citet{Oasa06}, and have used them to compare with the recent
evolutionary tracks of \cite{Baraffe15} that predict the absolute magnitudes in each band for stars of a given mass and age. The positions of these stars in the $(H-K_S), (J-H)$ diagram shows no evidence for infrared excess for any of them. Therefore, we have displaced their position in a three-dimensional diagram having the absolute $J$, $H$ and $K_S$ magnitudes as their axes along the vector representing the extinction law of \citet{Cardelli89}, until finding the closest point to the locus defined by the 0.5~Myr isochrone, thus obtaining the best-fitting mass, luminosity, and amount of foreground extinction. The 0.5~Myr isochrone has been chosen given the evidence discussed by \citet{Oasa06} for a cluster age below 1~Myr and the extreme youth of S106IR itself. The loci defined by isochrones corresponding to different ages are very similar, but the mass corresponding to a given absolute magnitude varies considerably with the age. Our results are therefore little sensitive to the chosen age regarding the best-fitting luminosity $L_{\rm bol}$ and foreground extinction $A_V$, but the masses derived should be increased if an older isochrone is chosen instead. Table~\ref{othervars_pars} shows the results of our best fits to the mass and foreground extinction of the four variable stars.

  A property common to all four stars is their high level of X-ray emission. A comparison with the results of \citet{Giardino04} shows that these four variable
stars include the 3rd, 4th, 6th and 20th most intense sources in their list of 87 X-ray emitters observed with {\it Chandra} in the S106 cluster, despite their inconspicuousness at infrared wavelengths. We have re-derived the X-ray luminosities of these sources in the $0.1-7.5$~keV range taking the observed X-ray fluxes reported by \citet{Giardino04}, now using the distance of 1.3~kpc adopted throughout our study, and the values of the foreground extinction obtained by us. The values thus obtained are also listed in Table~\ref{othervars_pars}, together with the derived $L_X / L_{\rm bol}$ ratios. The values are high, but well within the range found in similarly young populations like the Orion Nebula Cluster \citep[e.g.][]{Nunez16} for stars of approximately solar mass, in agreement with the masses that we estimate. These stars have convective envelopes in which the magnetic activity causing the X-ray emission also generates large spots in the stellar surface, which produce variability as they evolve and stellar rotation brings them in and out of sight. The variability timescales seen in Figure~\ref{othervars_J} are consistent with rotational modulation (see \citet{Fritzewski16} and \citet{Coker16} for recent work and references). Given the common characteristics of non-regular variability (except perhaps for J20271921+3722361), strong X-ray emission, and lack of significant infrared excess indicating the absence of large amounts of warm circumstellar matter, we consider starspots, rather than other causes like binarity or accretion, as the most likely cause of variability of the four objects presented here.

\end{appendix}





\end{document}